\documentclass{iau} 
\usepackage{graphicx}

\title[O abundance radial gradient evolution in MWG] 
{The  evolution of the Oxygen abundance radial gradient in the Milky Way Galaxy disk}

\author[Moll{\'a} et al.]   
{Mercedes Moll{\'a}$^1$, Oscar Cavichia$^{2}$, Roberto D. D. Costa$^{3}$, Walter  J. Maciel$^{3}$, Brad Gibson$^{4}$, 
\and Angeles I D{\'\i}az$^{5,6}$}

\affiliation{$^1$ CIEMAT, Avda. Complutense 40, 28040 Madrid, Spain
 \\ email: {\tt mercedes.molla@ciemat.es} \\[\affilskip]
$^2$Instituto de F\'{i}sica e Qu\'{i}mica, Universidade Federal de Itajub\'{a}, Av. BPS, 1303, 37500-903, Itajub\'{a}-MG, Brazil\\
$^{3}$ Instituto de Astronomia, Geofisica e Ci{\^e}ncias Atmosf{\' e}ricas, Universidade de S{\^a}o Paulo, 05508-900, S{\^a}o Paulo-SP, Brazil\\
$^{4}$ E.A Milne Centre for Astrophysics, University of Hull, HU6~7RX, 
United Kingdom\\
$^{5}$ Universidad Aut\'{o}noma de Madrid, 28049, Madrid, Spain \\
$^{6}$ Astro-UAM, Unidad Asociada CSIC, Universidad Aut{\'o}noma de Madrid,28049, Madrid, Spain
}

\pubyear{2016}
\volume{323}  
\jname{Planetary nebulae: Multi-wavelength probes of stellar and galactic evolution}
\editors{X. Liu, L. Stanghellini, \& A. Karakas, eds.}
\begin{document}

\maketitle

\begin{abstract}
We review the state of our chemical evolution models for spiral and low mass galaxies. We analyze the consequences of using different stellar yields, infall rate laws and star formation prescriptions in the time/redshift evolution of the radial distributions of  abundances, and other quantities as star formation rate or gas densities, in the Milky Way Galaxy; In particular we will study the evolution of the Oxygen abundance radial gradient analyzing its relation with the ratio SFR/infall. We also compare the results with  our old chemical evolution models, cosmological simulations and with the existing data, mainly with the planetary nebulae abundances.

\end{abstract}

\firstsection 
\section{Introduction}

Chemical elements are mostly created  by stellar nucleosynthesis, and eventually ejected and diluted in the interstellar medium when stars die. They
are then incorporated into the successive generations of stars. This cycle of formation and death of stars and the subsequent ejection of elements may occur many times, if the star formation rate (SFR) is high, with  elemental abundances increasing rapidly in the interstellar medium (ISM); If the SFR is low, the abundances will increase slowly, maintaining low during a long time. This way, elemental abundances give clues about when, how and with which rate stars form, existing a direct link between the evolutionary histories of galaxies or galaxy regions and the abundances measured within them. Chemical evolution models try to understand
the metal enrichment process, predicting the elemental abundances expected in each region as depending on different inputs
and hypotheses, as the Initial Mass Function (IMF), the star formation law, the existence or not of inflows or outflows of gas, or the
stellar nucleosynthesis yields. Once computed a model for a galaxy, it is possible to compare with the observational data
and deduce if  our assumptions are valid or it is necessary to modify them. 

Many numerical chemical evolution models have been presented in the literature from decades ago \cite{lf83,lf85}, \cite{dt84},\cite{td85}, \cite{mf89}, \cite{fer94}. All of them are usually computed to reproduce the Milky Way Galaxy (MWG) and other nearby spiral galaxies data as radial distributions of star formation, gas density, surface density stellar profile and the radial gradient of most common elements (C, N, O, Fe...), which define the present state of our Galaxy. The existence of a radial gradient of abundances in the spiral disks is well known \cite{hw99}. This gradient  was first observed in the MWG \cite{sha83} and later in other 
external galaxies \cite{mccall85,zar94}, and it is now well  characterized in our local Universe as shown by \cite{sanchez14}. However, even fitting the present day data, 
differences among models appear for other evolutionary times. In particular, the time evolution of the radial gradient of abundances is not the same for all models. Thus, in \cite{dt84,td85,mol90}  or \cite{chia97}  a flat radial gradient appears in the early times of evolution of a galaxy, steepening with time; on the contrary,   in \cite{fer94,pran95,mol97,por00}, and \cite{hou00},  the gradient is  steep, and then it flattens with time. In the first case, a disk is already formed  in time zero with an infall of primordial gas which dilutes the enrichment of the disk due to the stars death. In the second type of models, the disk forms out from the gas infalling onto the equatorial plane. Therefore,  the evolution of the metallicity gradient seems to be directly related on the formation process of the disk, stressing the importance of this point.

In \cite{mol97} we analyzed the existing data for MWG about this question:
planetary nebulae (PN) abundances, globular and open clusters metallicities and stellar abundances. Each data set has his own 
problems to determine the time evolution of the radial gradient. The PN are useful to estimate the radial gradient of Oxygen, an element
not modified by the nucleosynthesis in the progenitors stars of PN. The early data analyzing this question \cite{mac03} showed a radial gradient
steeper than the one for the present time, although, due to error bars, the observational points mixed with the H{\sc ii} regions data in 
the plot of O/H {\sl vs} galactocentric radius $R$. The open clusters problems are derived of the necessary classification in thin or thick disk or even halo 
populations before their use in a radial distribution of metallicity. Furthermore, the ages determination, coming from
a fit of  their spectra to stellar models, which depend on metallicity and on age simultaneously, added to differences depending on the use of LTE or NLTE models, may be uncertain. 
The estimate of age and metallicity of globular clusters (GC) seems less problematic since there was assumed that they form as a single stellar population, with a single burst of star formation and, consequently, to obtain only an age and a metallicity for all the stars of a GC\footnote{Recent
works conclude that there are at least two stellar generations in these old objects.} is more precise.  Based on these GCs data, 
we concluded that the radial gradient of abundances in MWG was steeper in the past, and it has been flattening with time
until arriving to the present value. However, these old objects may also be part of the thick disk or the halo populations.
As for the open clusters, the kinematic information is essential to determine their group and, therefore, to know if the corresponding data may be included in the study of the time evolution of the radial distribution of {\bf disk} abundances.

In \cite{md05}(hereinafter MD05), we show 440 chemical evolution models,
for 44 different galaxy masses and 10  possible star formation efficiencies in each one.
From a MWG-like model, we obtained the  evolution of the radial gradient of Oxygen abundances, finding
a gradient of $\sim -0.2$ dex\,kpc$^{-1}$ for $z=2$ which flattens with time  until reaching the present time value 
$\sim -0.05-0.06$ dex\,kpc$^{-1}$. This behavior was in agreement with the cosmological simulations from \cite{pil12}
and also with the \cite{yuan11} value found at $z\sim1.5$.  However, it is necessary to take into account that this gradient
was calculated with the O abundances of all radial regions computed in our model (which arrived until 25\,kpc). In fact, the optical
radius of MWG is $\sim 13-15$\,kpc, and, therefore, to include the regions out of a given radius might be considered as
inadequate, since there thick or halo populations may being mixed with the thin disk. Taking into account that these outer regions
of the galaxy are not evolved, this implies a steepening of the radial gradient, more strongly for the highest redshift since 
the parts not evolved of the disk will be more inner.  If we only consider the regions within the optical radius to compute
the radial gradient, the evolution of  this last one along the redshift is smoother, with a value  $\sim -0.06$ dex\,kpc$^{-1}$ for all redshifts.

We have now revised our chemical evolution models to include more realistic and updated inputs in our code. In this work we
analyze the corresponding results for our new MWG model, used to calibrating the model grid. We study in particular
the evolution of the radial gradient of Oxygen, searching for the possible differences with the old model from MD05 and checking 
if the new prescriptions related on stellar yields,  star formation or infall rate have modified the  behavior of this gradient. 
We also compare the results with the most recent data in the literature.

\section{Abundances in the MWG}

In the recent  years, new sets of data have been published, as consequence of better instruments, techniques  and telescopes.
\cite{anders16} give a summary of the observations (see their Fig.5), referring to PN, open clusters, and stars from MWG in order to analyze the evolution of the radial gradient of metallicity along the time.  The problem is that each type of objects shows a different evolution. Thus, PNs which gave a steep radial gradient for a time 
$\sim$8\,Gyr compared with the one corresponding to the present time given by H{\sc ii} regions, give now, following the most recent findings, a flat evolution along the time; that is, the radial gradient for O is basically the same from 5-6\,Gyr ago until now \cite{mac13}, or even has steepen slightly on time \cite{stan10}. A radial gradient constant along the time is also obtained by \cite{mag16} for M33, M31, M81 and NGC300,  comparing their radial distributions of Oxygen abundances given by H{\sc ii} regions and PN, finding no evidences of  evolution of the radial gradient with time. The PN give precise measurements of abundances since they are very bright, allowing to observe them even being far away. However, to estimate their masses (an in consequence their ages) and their distances is not an easy task, as explained in \cite{cav11},  which makes the Magrini et al. (2016) constraint not totally sure. From open clusters data, practically all of works,  as
 \cite{friel02, chen03,mag09,frin13,cun16}, arrive to a similar conclusion: the radial gradient of metallicity was steeper in the past than now. The number of points of the complete sample  in these works is usually small and, therefore, each defined bin of age is wide with a handful of points. Finally, stellar abundances,  as given \cite{nord04,casa11,berg14,xiang15}, show a flattening of the radial gradient. However, the oldest age bin (around 12-13\,Gyr old) has a radial gradient zero. It means that the radial gradient started flat, then it did steeper, and after it flattened again. \cite{anders16} do not agree with these results, obtaining a radial gradient steepening continuously with time. Looking at these data in detail, we find some problems for the precise determination of the radial gradient for
the old ages bin.  \cite{berg14} give the metallicity radial distribution as the Iron abundance dividing their sample in different stellar ages, with a radial range for the whole sample of $\sim 4$\,kpc (actually not very wide), but  the oldest stars are located in a reduced region around the Solar vicinity, with a radial range of $\sim 1$\,kpc. 
It is impossible to estimate a gradient with  such a short radial range. This problem for the oldest age bin also occurs in \cite{xiang15}, where the radial range reduces  from the 3-4\,kpc for the youngest bin to 1-1.5\,kpc for the oldest one. For the CoRoGee sample from  \cite{anders16} the total sample has a good radial range from 4\,kpc to 14\,kpc, but  the oldest bin has a range much shorter, from 8 to 12\,kpc. This produces a cloud of points around 
$\sim 9$\,kpc of galactocentric distance, which does not allow the computation of a precise radial gradient.

\begin{figure}[!ht]
\begin{center}
\includegraphics[width=3.7in,angle=-90]{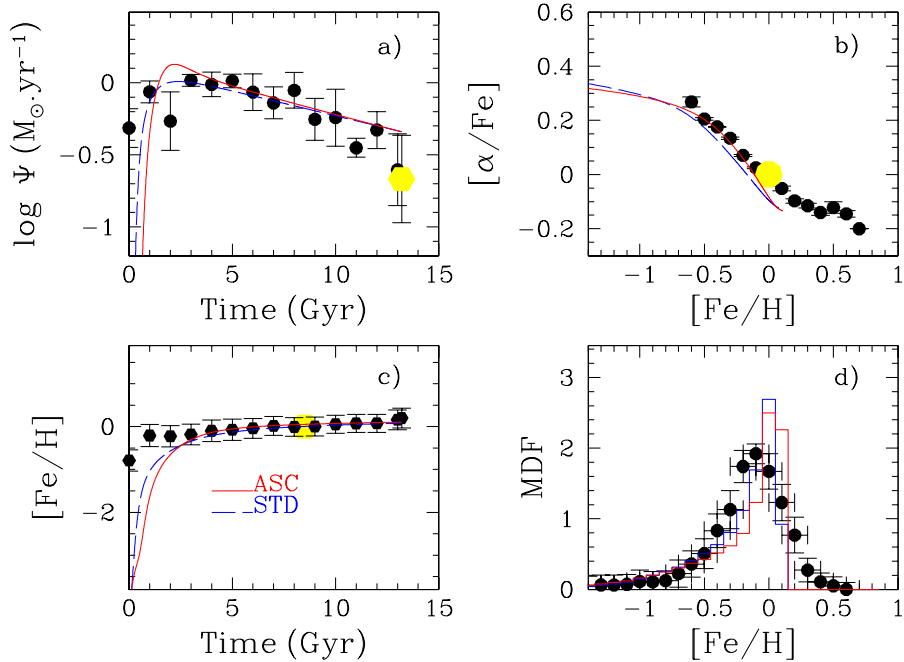} 
\caption{Solar Region evolution: a) SFR as a function of time; b) $[\alpha/Fe]$ as a function of $[Fe/H]$; c) 
$[Fe/H]$ as a function of time; and d) The metallicity distribution function. In all cases the blue and red lines represent the STD and ASC models,  
as labelled, while the black dots represent the
Solar Region data as compiled in \cite{mol15}(Appendix). The  yellow large dot is the value of the SFR at the present time in a), the Solar abundance in b) and c), located in the time when the Sun formed in this last panel.}
\label{fig1}
\end{center}
\end{figure}

\section{Multiphase Chemical Evolution models}

Our basic model is well described in \cite{fer94,md05}. We start with a spherical protogalaxy with a dynamical mass distribution 
given by the rotation curve. This initial gas falls onto the equatorial plane forming out the disk. In the halo the star formation 
follows a Schmidt-Kennicutt law, while in the disk molecular clouds form first, and then stars are created from cloud-cloud collisions and from the interaction of massive stars with the surrounding clouds. Including the star mean lifetime 
and the stellar yields, we compute abundances for 15 elements until Fe, being 
this element mostly created by the Supernova type Ia  yields from \cite{iwa99}. 
In this standard model (STD), the star and molecular cloud formation are calculated with two efficiencies
for each galaxy model,  which we vary simultaneously with values among 0 and 1.

\begin{figure}[!ht]
\begin{center}
\includegraphics[width=3.4in,angle=-90]{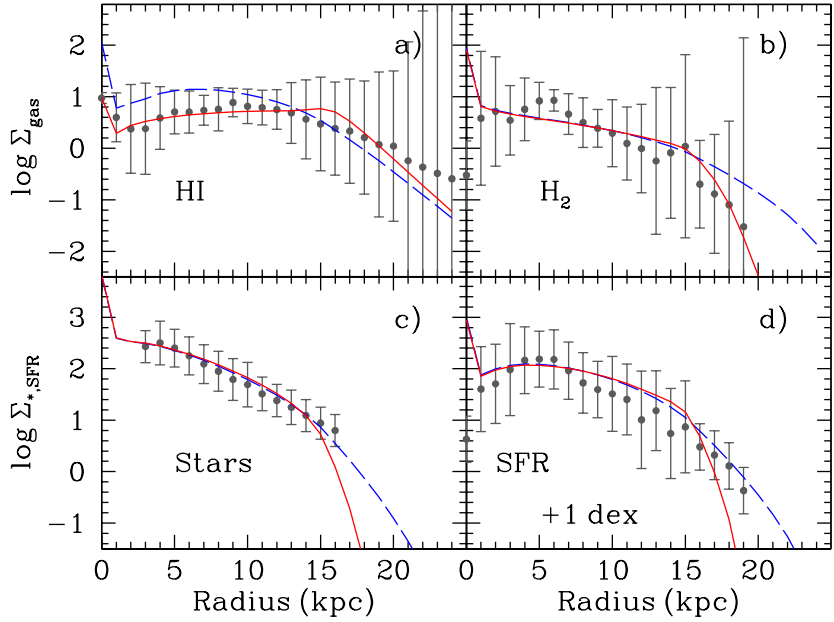} 
\caption{Radial distributions of surface density for: a) Diffuse gas; b) Molecular gas; c) Stellar profile;
and d) SFR. In all cases the blue and red lines represent the STD and ASC models while the black dots represent the
MWG data as compiled in the Appendix from  \cite{mol15}.}
\label{fig2}
\end{center}
\end{figure}

We have now an updated grid of models (Moll{\'a} et al, in preparation), where we use the stellar yields from \cite{gav05,gav06,lim03,chi04} and IMF from \cite{kro02}, selected as the best combination of stellar yields and IMF from 144 models in \cite{mol15}. We have also revised the infall rates of gas, calculated to create disks as observed following \cite{sal07}. Details are given in \cite{mol16}. 
Moreover, we now follow the prescription by Ascasibar et al. (in preparation) to form molecular clouds from the diffuse gas. This model (ASC)
is compared with our standard model (STD). The details are given in \cite{mol17}.

For a MWG-like model, we use a dynamical mass of 10$^{12}$\,M$_{\odot}$, able to create a disk of 
$\sim 7\,10^{10}$\,M$_{\odot}$. In Fig.~\ref{fig1} we represent the evolution of the region located a $R=8$\,kpc 
as function of time in panels a) and c), and as function of the metallicity [Fe/H] in panels b) and d), compared with the
Solar Region data.  In  Fig.~\ref{fig2} we draw the surface densities of HI, H$_{2}$, stars and SFR for the disk 
component, compared  with the MWG data. We compare both STD and ASC models, finding that basically both
fit the data within errors, with differences only in the outer regions with a more extended disk for STD.

\section{Results: the time evolution of Oxygen radial gradient}

\begin{figure}[!ht]
\begin{center}
\includegraphics[width=3.4in,angle=-90]{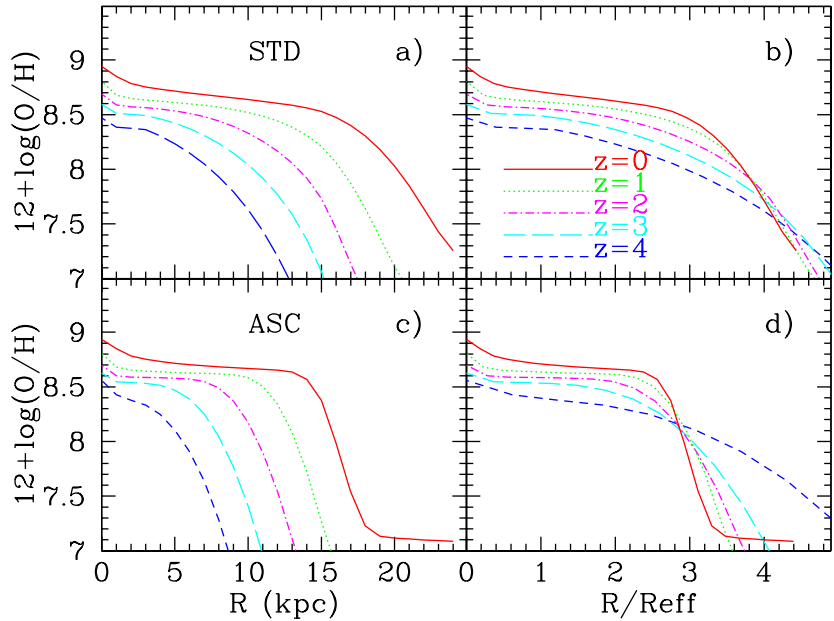} 
\caption{Radial distributions of the $\rm 12+log(O/H)$ abundances for different redshifts as labelled. Left panels as function of the galactocentric distance (in kpc) and right panels as
function of the normalized radius $R/R_{eff}$.  Top panels represent the evolution of STD, while the bottom ones refer to ASC.}
\label{fig3}
\end{center}
\end{figure}

In Fig.~\ref{fig3} we show the radial distribution for O/H as given by STD (top panels) and ASC (bottom panels) prescriptions, for different
redshifts as labelled in the top right panel. This way, we may check the effect of the star formation law on the evolution  of the
radial gradient.  The distributions shows a sharp cut of the disk in ASC while this is smoother in STD, with a more extended disk, as said before.
This is due to the effect of threshold of  ASC, since the necessary density or metallicity must be reached to create molecular clouds.  In left panels we show these distributions as a function of the galactocentric distance $R$ (in kpc). In right panels we show the same
distributions as a function of the normalized radius $R/R_{eff}$.   
It is clear that the slope of the distribution within the optical disk, defined as $R\sim 3\,R_{eff}$ ($\sim 15$\,kpc for the STD model at $z=0$),
is almost the same for all redshifts, as shown in right panels. We must remember that  $R_{eff}$ changes with redshift:  
the stellar disk increases their size when stars form in an inside-out scenario; at the early times stars form only in the inner disk. In the STD  model, $R_{eff}$ is around 5\,kpc for the present time, but it is $\sim 2.5$\,kpc at $z=4$. However, $R_{eff}$ 
increases very slowly in ASC model, being $R_{eff}\sim 1.6$\,kpc at $z=4$.  Thus, at this redshift, the optical radius would be around $\sim 7$ \,kpc in STD model,
but it only arrives to $\sim 4.5$\,kpc in ASC. Therefore, it seems quite evident that the slope of the radial gradient
is basically the same for all redshifts when is measured within the optical radius, where we must define a radial gradient of the disk oxygen abundance.  
If a radial gradient different than the one of the present time is obtained at other time/redshift, it implies that it 
is being measured with regions out of the optical disk, with stellar populations of the thick disk, or of the halo, not unexpected if we are observing the phases of the disk formation at $z>2.5$.

\begin{figure}[t]
\begin{center}
\includegraphics[width=3.4in,angle=-90]{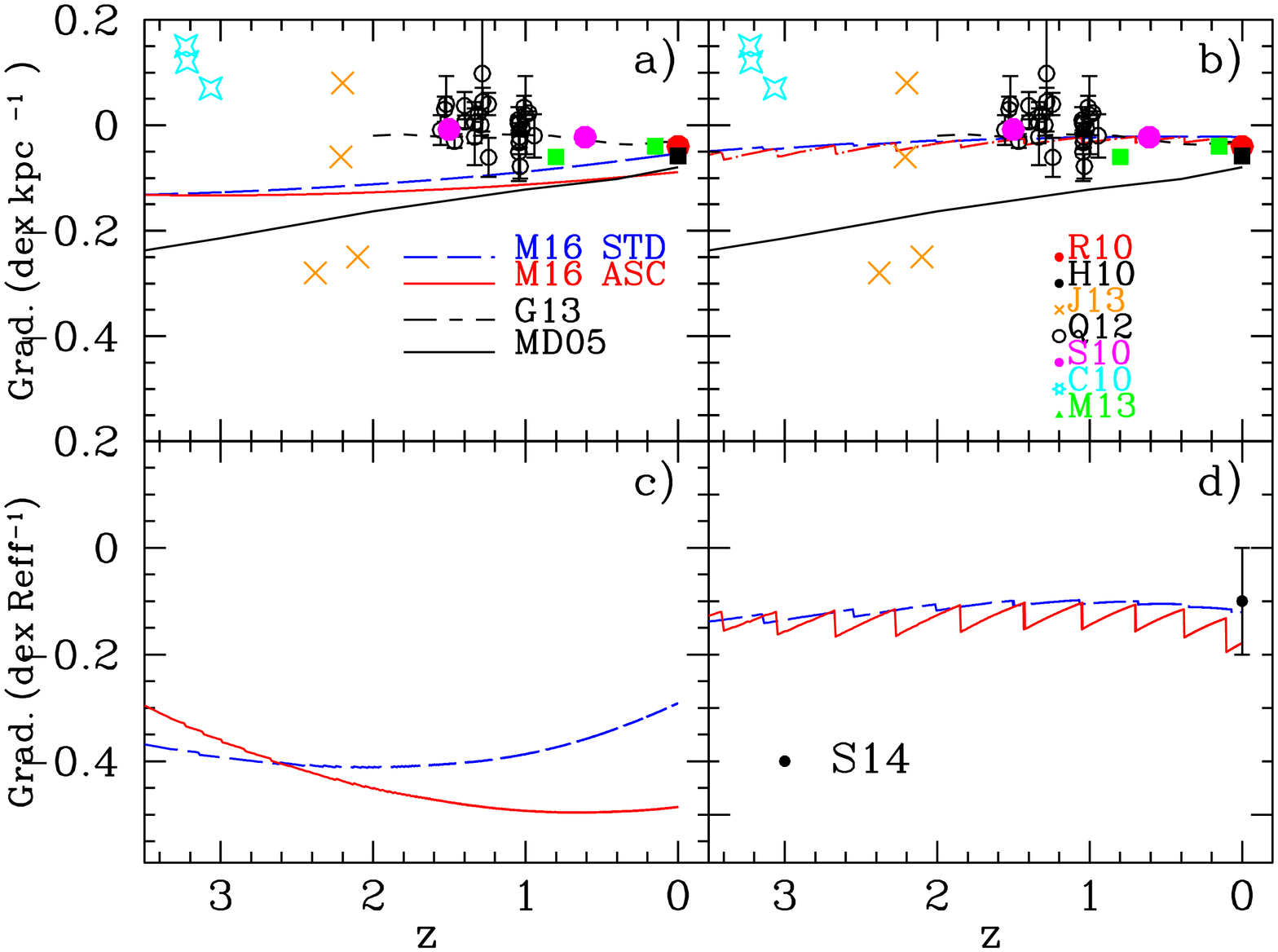} 
\caption{Evolution of the O/H radial gradient with the redshift $z$ in dex\,kpc$^{-1}$ (top panels) and dex\,$R_{eff}^{-1}$ (bottom panels). In a) and c) calculated with the whole radial range. In b) and d) calculated with $R\le 3\,R_{eff}$. In top panels, data are from \cite{hen10}(H10), \cite{rup10}(R10), \cite{jon13}(J13), \cite{quey12}(Q12), \cite{stan10}(S10), \cite{cresci10}(C10) and \cite{mac13}(M13). Models STD and ASC  are our new models (M16) from \cite{mol17}, our old model (MD05) and \cite{gib13}(G13). In panel d) the present day value corresponds to the average gradient found by \cite{sanchez14}(S14).}
\label{fig4}
\end{center}
\end{figure}

We summarize our results in Fig.~\ref{fig4}, with the evolution along redshift of the O/H radial gradient, measured as dex\,kpc$^{-1}$ in the top panels
and as dex\,R$_{eff}^{-1}$ in the bottom ones. In left panels  the gradient is calculated using the whole radial range that we run our MWG model, while in the right ones we use only regions within the optical radius. Top panel results are compared with our old model MD05, cosmological simulations by \cite{gib13} and recent data as labelled, including \cite{jon13} and \cite{cresci10} for high redshift galaxies.

\section{Conclusions}
The main conclusions can be summarised as:
\begin{enumerate}
\item The radial range where the least-squares straight line fit is done  is crucial to determine the value of the abundance radial gradient.
\item Our MWG models produce a smooth evolution of the O abundances radial gradient  with a value of  -0.15\,dex\,kpc$^{-1}$ at $z=3.5$ arriving
to -0.05  dex\,kpc$^{-1}$ at $z=0$ when the whole radial range is used.
\item The radial gradient has a value around $\sim -0.05$\,dex\,kpc$^{-1}$  for $z \le 3.5$ when it is measured within the optical radius (variable with $z$).
 \item The radial gradient computed by our new models is flatter than the obtained in MD05 as consequence of the new infall rates, smoother than the old ones.
\item If the radial gradient is measured as a function of a normalized radius ($R/R_{eff}$), differences arise between STD and ASC proceeding from the different
grow of size of the stellar disk in both models (even with the same infall rates), if all radial regions are used.
\item If the radial gradient is computed within the optical radius, as a function of a normalized radius ($R/R_{eff}$), it has a constant value of
$\sim  -0.15$ \,dex\,R$_{eff}^{-1}$ for  $z \le 3.5$, which agrees with \cite{sanchez14} result for the CALIFA local galaxies.
\end{enumerate}


\begin{thebibliography}{}

\bibitem[Anders, Chiappini \& Minchev (2016)]{anders16} 
Anders, F., Chiappini, C., Minchev, I., et al.\ 2016,  A\&A, submitted (arXiv:1608.04951) 

\bibitem[Bergemann et al. (2014)]{berg14} 
Bergemann M., et al., 2014, \textit{A\&A}, 565, A89 

\bibitem[Casagrande et al. (2011)]{casa11} 
Casagrande, L., Sch{\"o}nrich R., Asplund M., Cassisi S., et al., 2011, \textit{A\&A}, 530, A138 

\bibitem[Cavichia, Costa, \& Maciel (20131)]{cav11}
Cavichia O., Costa R.~D.~D., Maciel W.~J., 2011, \textit{RMxAA}, 47, 49 

\bibitem[Chen, Hou, \& Wang (2003)]{chen03} 
Chen L., Hou J.~L., Wang J.~J., 2003, \textit{AJ}, 125, 1397 

\bibitem[Chiappini, Matteucci \& Gratton (1997)]{chia97}
Chiappini, C., Matteucci, F., \& Gratton, R.\ 1997, \textit{ApJ}, 477, 765

\bibitem[Chieffi \& Limongi (2004)]{chi04}
Chieffi A., Limongi M., 2004, \textit{ApJ}, 608, 405

\bibitem[Cresci et al. (2010)]{cresci10}
Cresci, G., Mannucci, F., Maiolino, R., et al.\ 2010, \textit{Nature}, 467, 811

\bibitem[Costa, Cavichia, \& Maciel (2013)]{cavichia13} 
Costa, R.~D.~D., Cavichia, O., Maciel, W.~J., 2013, \textit{IAUS}, 289, 375 

\bibitem[Cunha et al. (2016)]{cun16}
Cunha, K., Frinchaboy, P.~M., Souto, D., et al.\ 2016, \textit{Astronomische Nachrichten}, 337, 922

\bibitem[D{\'{\i}}az \& Tosi (1984)]{dt84}
Diaz, A.~I., Tosi, M., 1984, \textit{MNRAS}, 208, 365

\bibitem[Ferrini et al. (1994)]{fer94}
Ferrini, F., Molla, M., Pardi, M.~C., Diaz, A.~I., 1994, \textit{ApJ}, 427, 745

\bibitem[Friel et al. (2002)]{friel02} 
Friel, E.~D., Janes, K.~A., Tavarez, M., Scott, J., Katsanis, R., et al., 2002, \textit{AJ}, 124, 2693 

\bibitem[Frinchaboy et al. (2013)]{frin13} 
Frinchaboy, P.~M., et al., 2013, \textit{ApJ}, 777, L1 

\bibitem[Gavil\'an, Buell \& Moll\'a (2005)]{gav05}
Gavil{\'a}n, M., Buell, J.~F., \& Moll{\'a}, M., 2005, \textit{A\&A}, 432, 861

\bibitem[Gavil\'an, Moll\'a \& Buell (2006)]{gav06}
Gavil{\'a}n, M., Moll{\'a}, M., \& Buell, J.~F., 2006, \textit{A\&A}, 450, 509

\bibitem[Gibson et al. (2013)]{gib13}
Gibson, B.~K., Pilkington, K., Brook, C.~B., Stinson, G.~S., \& Bailin, J.\ 2013, \textit{A\&A}, 554, A47

\bibitem[Henry \& Worthey (1999)]{hw99}
Henry, R.~B.~C., \& Worthey, G.\ 1999, \textit{PASP}, 111, 919

\bibitem[Henry et al. (2010)]{hen10}
Henry, R.~B.~C., Kwitter, K.~B., Jaskot, A.~E., et al.\ 2010, \textit{ApJ}, 724, 748

\bibitem[Hou, Prantzos \& Boissier (2000)]{hou00}
Hou, J.~L., Prantzos, N., \& Boissier, S.\ 2000, \textit{A\&A}, 362, 921 

\bibitem[Iwamoto et al. (1999)]{iwa99}
Iwamoto, K., Brachwitz, F., Nomoto, K., et al.\ 1999, \textit{ApJS}, 125, 439

\bibitem[Jones et al. (2013)]{jon13}
Jones, T., Ellis, R.~S., Richard, J., \& Jullo, E.\ 2013, \textit{ApJ}, 765, 48 


\bibitem[Kroupa (2002)]{kro02}
Kroupa, P., 2002, \textit{Sci}, 295, 82
 
\bibitem[Lacey \& Fall (1983)]{lf83}
Lacey, C.~G., \& Fall, S.~M.\ 1983, \textit{MNRAS}, 204, 791

\bibitem[Lacey \& Fall (1985)]{lf85}
Lacey, C.~G., \& Fall, S.~M.\ 1985, \textit{ApJ}, 290, 154 

\bibitem[Limongi \& Chieffi (2003)]{lim03}
Limongi, M., \& Chieffi, A., 2003, \textit{ApJ}, 592, 404

\bibitem[Maciel, Costa \& Uchida (2003)]{mac03}
Maciel, W.~J., Costa, R.~D.~D., \& Uchida, M.~M.~M., 2003, \textit{A\&A}, 397, 667 

\bibitem[Maciel \& Costa (2013)]{mac13}
Maciel, W.~J., \& Costa, R.~D.~D.\ 2013, \textit{RMxA\&Ap}, 49, 333

\bibitem[Magrini et al. (2009)]{mag09}
Magrini, L., Sestito, P., Randich, S., \& Galli, D.\ 2009, \textit{A\&A}, 494, 95 

\bibitem[Magrini et al. (2016)]{mag16}
Magrini, L., Coccato, L., Stanghellini, L., Casasola, V., Galli, D.\, 2016, \textit{A\&A}, 588, A91 

\bibitem[Matteucci \& Francois (1989)]{mf89}
Matteucci, F., \& Francois, P.\ ,1989, \textit{MNRAS}, 239, 885
 
\bibitem[McCall, Rybski \& Shields (1985)]{mccall85}
McCall, M.~L., Rybski, P.~M., \& Shields, G.~A.\,1985, \textit{ApJS}, 57, 1

\bibitem[Moll{\'a}, D{\'\i}az \& Tosi (1990)]{mol90}
Moll{\'a}, M., D{\'\i}az, A.I., Tosi, M., 1990,  in Chemical and dynamical evolution of Galaxies, 
eds. F. Ferrini, J. Franco, \& F. Matteucci, ETS Editrice (PISA), 577

\bibitem[Moll{\'a}, Ferrini \& Diaz (1997)]{mol97}
Moll{\'a}, M., Ferrini, F., \& Diaz, A.~I.\ 1997, \textit{ApJ}, 475, 519

\bibitem[Moll{\'a} \& D{\'{\i}}az (2005)] {md05}
Moll{\'a}, M. \& D{\'{\i}}az, A.I.\ 2005, \textit{MNRAS}, 358, 521 (MD05)

\bibitem[Moll{\'a} et al. (2015)]{mol15}
Moll{\'a}, M., Cavichia O., Gavil{\'a}n M., \& Gibson B.~K.\ 2015, \textit{MNRAS}, 451, 3693

\bibitem[Moll{\'a} et al. (2016)]{mol16}
Moll{\'a}, M., D\'{\i}az, A.~I., Gibson, B.~K.,  et al., 2016, \textit{MNRAS}, 462, 1329

\bibitem[Moll{\'a} et al. (2017)]{mol17}
Moll{\'a}, M., Ascasibar, Y., D\'{\i}az, A.~I., \& Gibson, B.~K., 2017, \textit{MNRAS}, to be submitted

\bibitem[Nordstr{\"o}m et al. (2004)]{nord04} 
Nordstr{\"o}m B., et al., 2004, A\&A, 418, 989 

\bibitem[Pilkington et al. (2012)]{pil12}
Pilkington, K., Few, C.~G., Gibson, B.~K., et al.\ 2012, \textit{A\&A}, 540, A56

\bibitem[Portinari \& Chiosi (2000)]{por00}
Portinari, L., \& Chiosi, C.\ 2000, \textit{A\&A}, 355, 929

\bibitem[Prantzos \& Aubert (1995)]{pran95}
Prantzos, N., \& Aubert, O.\ 1995, \textit{A\&A}, 302, 69

\bibitem[Queyrel et al. (2012)]{quey12}
Queyrel, J., Contini, T., Kissler-Patig, M., et al.\ 2012, \textit{A\&A}, 539, A93

\bibitem[Rupke, Kewley \& Chiel (2010)]{rup10}
Rupke, D.~S.~N., Kewley, L.~J., \& Chien, L.-H.\ 2010, ApJ, 723, 1255

\bibitem[Salucci et al. (2007)]{sal07}
Salucci, P., Lapi, A., Tonini, C., Gentile, G., et al.., 2007, \textit{MNRAS}, 378, 41

\bibitem[S{\'a}nchez et al. (2014)]{sanchez14}
S{\'a}nchez, S.~F., Rosales-Ortega, F.~F., Iglesias-P{\'a}ramo, J., et al.\ 2014, \textit{A\&A}, 563, A49

\bibitem[Shaver et al. (1983)]{sha83}
Shaver, P.~A., McGee, R.~X., Newton, L.~M., Danks, A.~C.,  et al.\ 1983, \textit{MNRAS}, 204, 53

\bibitem[Stanghellini \& Haywood (2010)]{stan10}
Stanghellini, L., \& Haywood, M.\ 2010, \textit{ApJ}, 714, 1096 

\bibitem[Tosi \& D{\'\i}az (1985)]{td85}
Tosi, M., \& D{\'\i}az, A.~I.\ 1985, \textit{MNRAS} 217, 571

\bibitem[Xiang et al. (2015)]{xiang15}
Xiang, M.-S., Liu, X.-W., Yuan, H.-B., et al.\ 2015, \textit{Research in Astronomy and Astrophysics}, 15, 1209 

\bibitem[Yuan et al. (2011)]{yuan11}
Yuan, T.-T., Kewley, L.~J., Swinbank, A.~M., et al. .\ 2011, \textit{ApJL}, 732, L14

\bibitem[Zaristky, Kennicutt \& Huchra (1994)]{zar94}
Zaritsky, D., Kennicutt, R.~C., Jr., \& Huchra, J.~P.\ 1994, \textit{ApJ}, 420, 87

\end{thebibliography}
\end{document}